\documentclass{article}\usepackage{emulateapj}
\usepackage{psfig}\usepackage{apjfonts}

\def\gtrapprox{\;\lower 0.5ex\hbox{$\buildrel >\over \sim\ $}}
\def\lessapprox{\;\lower 0.5ex\hbox{$\buildrel < \over \sim\ $}}

\def\phiI  {$\varphi_i$}

\def\Pcos  {$\Phi^0$}

\def\tauLL {\bar\tau}

\def\aB    {\alpha_{\scriptscriptstyle B}}

\def\HI    {H{$\rm\scriptstyle I$}}
\def\Ha    {${\rm H}\alpha$}
\def\eg    {{\it e.g.,\ }}
\def\ie    {{\it i.e.,\ }}
\def\cf    {{\it cf.\ }}
\def\qv	   {{\it q.v.,\ }}
\def\etal  {{\it\ et al.}}
\def\kms   {\ km s$^{-1}$}
\def\intensity{\ifmmode{{\rm erg\ cm}^{-2}{\rm\ s}^{-1}
      {\rm\ Hz}^{-1}{\rm\ sr}^{-1}}
      \else {erg cm$^{-2}$ s$^{-1}$ Hz$^{-1}$ sr$^{-1}$}\fi}
\def\Em{\ifmmode{{\cal E}_m}\else {{\cal E}$_m$}\fi}
\def\Dm{\ifmmode{{\cal D}_m}\else {{\cal D}$_m$}\fi}
\def\fesc{\ifmmode{f_{\rm esc}}\else {$f_{\rm esc}$}\fi}
\def\rsolar{\ifmmode{r_\odot}\else {$r_\odot$}\fi}
\def\emunit{\ifmmode{{\rm cm}^{-6}{\rm\ pc}}\else {
cm$^{-6}$ pc}\fi}
\def\flux{\ifmmode{{\rm erg\ cm}^{-2}{\rm\ s}^{-1}}\else {erg
cm$^{-2}$ s$^{-1}$}\fi}
\def\fluxdensity{\ifmmode{{\rm erg\ cm^{-2}\ s^{-1}\ Hz^{-1}}}\else {erg
cm$^{-2}$ s$^{-1}$ Hz$^{-1}$}\fi}
\def\phoflux{\ifmmode{{\rm phot\ cm}^{-2}{\rm\ s}^{-1}}\else {phot
cm$^{-2}$ s$^{-1}$}\fi}
\def\phorate{\ifmmode{{\rm phot\ s}^{-1}}\else {phot s$^{-1}$}\fi}

\begin{document}

\title{The Escape of Ionizing Photons from the Galaxy}

\author{J. Bland-Hawthorn}
\affil{Anglo-Australian Observatory, P.O. Box 296, Epping, NSW 2121, Australia}
\author{Philip R. Maloney}
\affil{CASA, University of Colorado, Boulder, CO 80309-0389}

\begin{abstract}
The Magellanic Stream and several high velocity clouds have now been
detected in optical line emission. The observed emission measures and
kinematics are most plausibly explained by photoionization due to hot,
young stars in the Galactic disk. The highly favorable orientation of
the Stream allows an unambiguous determination of the fraction of
ionizing photons \fesc\ which escape the disk. We have modelled the 
production and transport of ionizing photons through an opaque 
interstellar medium.  Normalization to the Stream detections requires
$\fesc\approx 6\%$, in reasonable agreement with the flux required to
ionize the Reynolds layer. Neither shock heating nor emission within a
hot Galactic corona can be important in producing the observed \Ha\
emission. If such a large escape fraction is typical of $L_*$
galaxies, star-forming systems dominate the extragalactic ionizing
background. Within the context of this model, both the
three-dimensional orientation of the Stream and the distances to
high-velocity clouds can be determined by sensitive \Ha\ observations.

\end{abstract}


\section{Introduction}
\nobreak 
There is extensive theoretical and observational interest in
establishing what fraction of the total ionizing luminosity from the
stellar disk of the Milky Way and other galaxies escapes into the halo
and the intergalactic medium (\eg Miller \& Cox 1993; Dove \& Shull
1994; Leitherer\etal\ 1996). This quantity is also of cosmological
interest, since if $\fesc$ exceeds a few per cent, then star-forming
galaxies dominate over AGN in producing the ionizing background at low
redshift (Giallongo, Fontana, \& Madau 1997). Observationally, it is
extremely difficult to measure this quantity directly. Leitherer\etal
(1995) used {\it HUT} observations of four UV-bright galaxies to
constrain \fesc\ to typically no more than a few percent at the Lyman
limit. However, Hurwitz, Jelinsky \& Dixon (1997) showed that
absorption by gas within the Milky Way, which was ignored by
Leitherer\etal, raises these limits substantially, to $\gtrapprox
10\%$, so that they are no longer significant constraints.

Recently, Weiner \& Williams (1996; hereafter W$^2$) have detected \Ha\ 
emission at
three points along the Magellanic Stream. We demonstrate that this 
signal is almost certainly due to ionizing photons
escaping from the disk of the Galaxy. In the following section, we
present a simple model for the production and escape of ionizing
photons from the disk. In \S 3, we use the \Ha\ observations to
measure the mean ionizing photon opacity ($\tauLL$) of the disk, which
determines both the shape and normalization of the Galactic halo
ionizing field. Finally, in \S 4, we discuss the implications for
observations of the Stream, high-velocity clouds, and the cosmic
ionizing background.

\section{Galactic photoionization model}
\nobreak
\noindent{\sl Galactic disk.} We construct an idealized model for the
ionizing radiation field away from the Galactic plane. The coordinate
system is shown in Fig.~1. The ionizing stars are assumed to
be isotropic emitters confined to a thin disk; since we are interested
in emitting regions at distances much greater than the stellar disk
thickness, the neglect of the finite thickness of the disk is
reasonable. At coordinates $(x_o,0,z_o)$, a distance $R_o$ from an
arbitrary patch of the disk $dA$, the received flux $f_d$ (in units of
\fluxdensity) from ionizing disk sources with specific intensity
$\zeta_\nu$, through a solid angle $d\Omega$, is
\begin{equation}
f_d = \int_\Omega \zeta_\nu\ {\rm d}\Omega = \int \zeta_\nu\ \cos\theta_o\
{dA(r_o,\phi_o)\over R_o^2}
\end{equation}
where $dA = r_o\ dr_o\ d\phi_o$ and $R_o^2 = x_o^2 + z_o^2 + r_o^2 - 2 x_o r_o
\cos\phi_o$.
The normally incident, ionizing photon flux from the disk (in units of
\phoflux) is
\begin{equation}
\varphi_d(r_o,\phi_o) = \int_A n_d(r_o,\phi_o)\ \cos\theta_o\ {dA(r_o,\phi_o)\over R_o^2} 
= \int {d\sigma(r_o,\phi_o) \cos\theta_o\over R_o^2}
\end{equation}
for which $n_d$ and d$\sigma$ are the surface photon density and
brightness, respectively, within each disk element $dA$.

Vacca\etal\ (1996) have compiled a list of 429 O stars within 2.5 kpc
of the Sun, from which they determine an ionizing photon surface density at
the Solar Circle of $n_d(\rsolar)\approx 3.74\times 10^7\,\phoflux$,
where $\rsolar\simeq 8.0\pm 0.5$ kpc (Reid 1993). We assume an
axisymmetric exponential disk with radial scale length $r_d$, so that
$n_d(r) = n_o e^{-r/r_d}$. With $r_d =$ 3.5 kpc (Kent, Dame \&
Fazio 1991), we thus derive $n_o\approx 3.7\times 10^8\,\phoflux$.   
The explicit contribution from B stars leads to only a small correction.
The O star survey region includes part of the Local Arm and extends to the 
neighboring Sagittarius-Carina and Perseus spiral arms
(see Taylor \& Cordes 1993). Observations with the {\it Ultraviolet 
Imaging Telescope} of face-on, mid to late-type spiral galaxies
(made available by E.P. Smith) give us confidence that our extrapolation 
is reasonable. All integrations are performed out to 25 kpc in
radius (\ie de Geus\etal\ 1993; Ferguson\etal\ 1998), giving an integrated 
photon flux over the disk of $2.6\times 10^{53}$ \phorate. This value is
consistent with Mezger's (1978) estimate and within 25\% of the rate
inferred from COBE/FIRAS observations of the Galactic [NII] 205$\mu$m
emission (Bennett\etal\ 1994).

\psfig{file=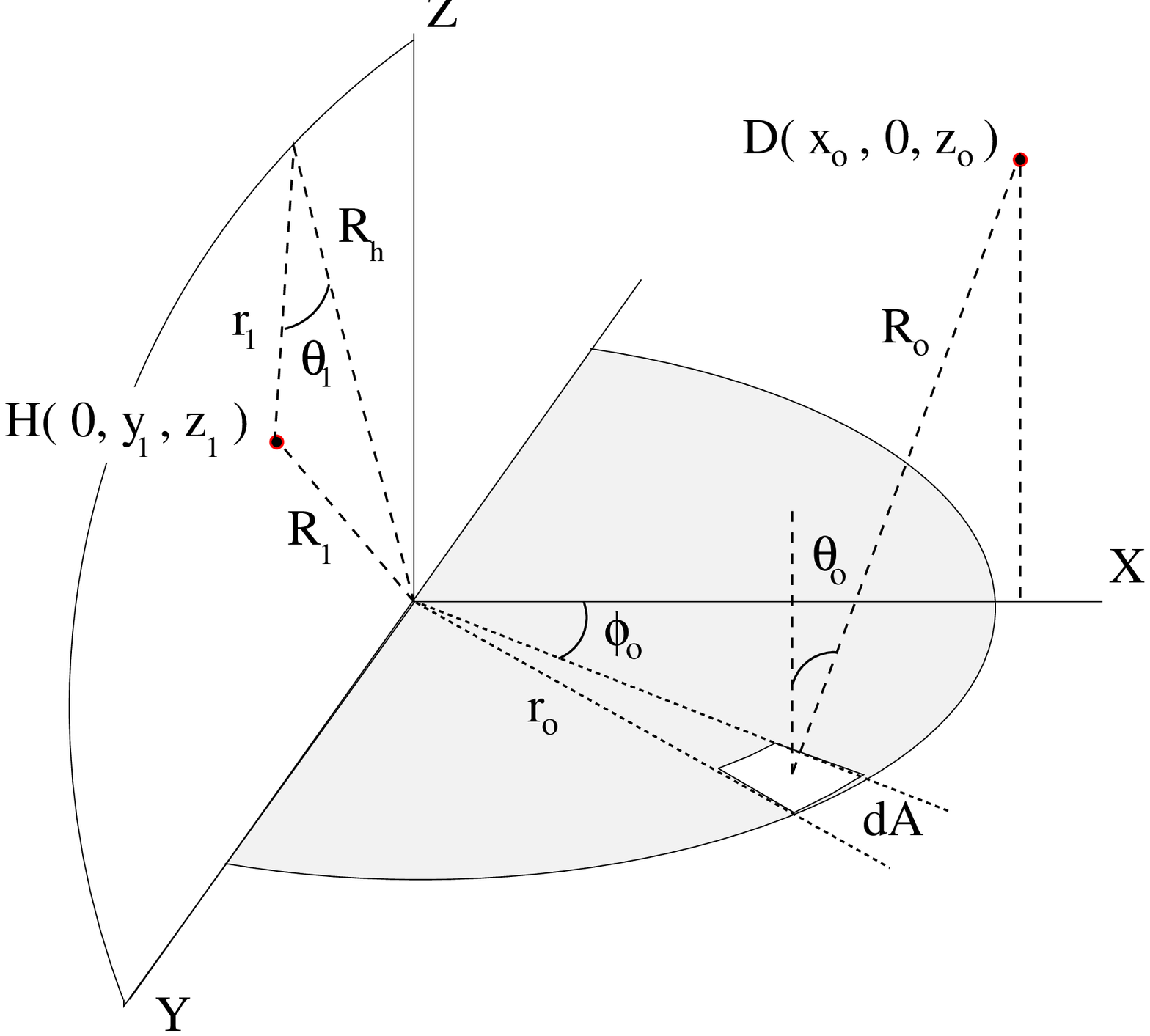,width=8.5cm,angle=-90}

\smallskip\noindent {\sl Fig.~1.}\ Coordinate system for a point $D(x_o,0,z_o)$ 
situated above an opaque disk, and a point $H(0,y_1,z_1)$ within an 
extended halo. The limit of integration $r_1$ for the halo is a function 
of $\theta_1$.

\bigskip
For an opaque disk, the patch $dA$ is observed through the intervening
disk interstellar medium (ISM) with optical depth $\tau$ (at a given
frequency), such that the observed surface brightness $d\sigma^\prime
= e^{-\tau(r_o,\phi_o)}\, d\sigma$. We approximate the disk ISM by a
smooth distribution. Since the absorbing disk thickness ($\sim$100 pc)
is negligible compared to the height of the Magellanic Stream above
the Galactic plane, we can assume plane-parallel absorption for which
$\tau(r_o,\phi_o) = \kappa R_o / z_o$, where $\kappa$ is the photon
absorption perpendicular to the galactic disk. Since for the purposes
of this paper we are only interested in the total ionizing photon
flux, and not the detailed spectral shape, we characterize the
radiation field by the mean ionizing photon opacity $\tauLL$ (\ie the
value averaged over photon energy).

\smallskip
\noindent{\sl Galactic corona.} Spitzer (1956) first suggested that
the Galaxy has a hot corona, with a scale height of several
kiloparsecs. Absorption line observations of highly ionized species
have confirmed the presence of hot gas at kiloparsec heights above the
Galactic plane (\eg Savage\etal\ 1989; Sembach \& Savage 1992). The
presence of an actual large-scale hot corona, as opposed to a patchy
distribution of hot gas arising in the disk, is more uncertain (\eg
Wang \& McCray 1993). To investigate the influence of such a corona,
we assume a distribution of hot gas of the form $n(R_1) =
n_c/(1+R_1^2/r_c^2)$ cm$^{-3}$, where $n_c$ is the core density and
$r_c$ the core radius, and a gas temperature $T_{\scriptscriptstyle h}$.

The halo is assumed to be finite in size with radius 
$R_{\scriptscriptstyle h}$. The normally incident flux on a (plane
parallel) cloud face, tangential to the radius vector, at a 
distance $R_1$ from the origin is described by
\begin{equation}
f_h = \int_\Omega {\xi_\nu \over r_1^2}\cos\theta_1\, d\Omega = 2\pi \int_o^{\pi/2}
F\left(\theta_1,{r_c\over R_1},{r_1\over R_1}\right)\ d\theta_1
\end{equation}
for which $\xi_\nu$ is the photon emissivity, $F$ is a somewhat
complicated function of $R_1$, $\theta_1$, $r_c$, and the
($\theta_1$-dependent) radial limit of integration $r_1$, which is given by
\begin{equation}
r_1 = {R_{\scriptscriptstyle h} \sin(\theta_1 + \sin^{-1} [R_1\sin\theta_1/R_{\scriptscriptstyle h}])\over\sin\theta_1} .
\end{equation}

\psfig{file=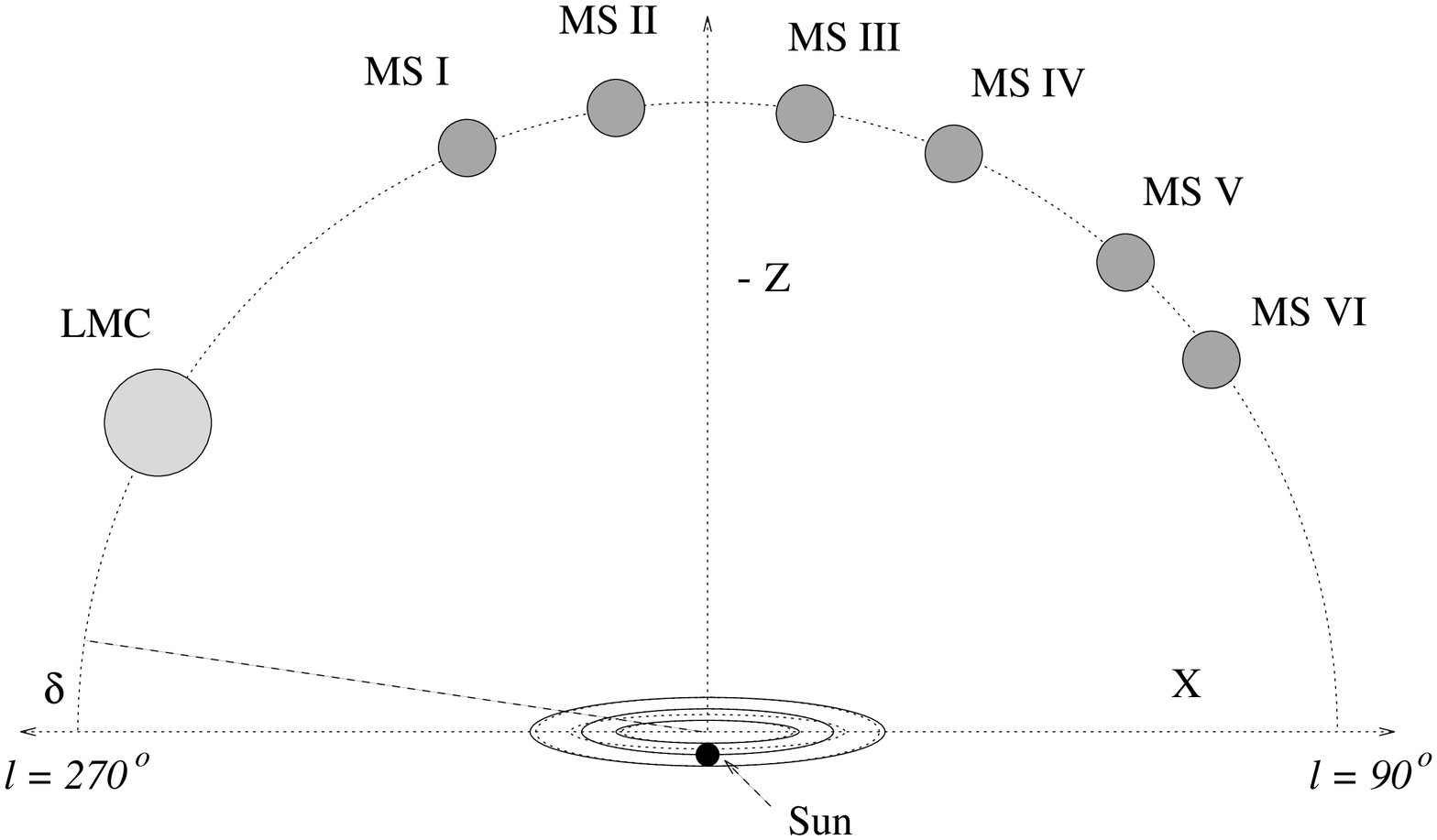,width=9cm}

\smallskip\noindent {\sl Fig.~2.}\ An illustration of the LMC and the dominant 
clouds in 
the Magellanic Stream (Mathewson \& Ford 1984) projected onto the Galactic 
$X$-$Z$ plane. The orbit of the Stream lies closer to the Great Circle whose 
longitude is $l = 285^\circ$.  We ignore small projection errors resulting 
from our vantage point at the Solar Circle. The angle $\delta$ is measured 
from the negative $X$ axis towards the negative $Z$ axis where 
$\delta = -b\ (0^\circ \leq \delta \leq 90^\circ)$ and 
$\delta = b+180^\circ\ (90^\circ \leq \delta \leq 180^\circ)$.

\bigskip
The photon emissivity $\xi_\nu$ was computed with the MAPPINGS 3 ionization
code (Sutherland \& Dopita 1996) and incorporated bremsstrahlung, line and 
non-equilibrium cooling.  The 
ionizing photon flux on the inner face of a cloud is given to 10\%
accuracy (for $R_1/r_c\le 12$) by
\begin{equation}
\label{phoflux}
\varphi_h(R_1) \approx 10^2 n^2_{-3} r_c {\xi_i\over 10^{-14}}
\left[{{\eta+1.3(R_1/r_c)^{1.35}}\over{(1+R_1^2/r_c^2)^{1.5}}}\right]\;
\phoflux
\end{equation}
where 10$^{-3}n_{-3}$ cm$^{-3}$ is the core number density, $R_1$ and
$r_c$ are in kpc, and $\xi_i$ is the frequency-integrated ionizing
photon emissivity. For an optically thin Galactic disk, $\eta = \pi/4$;
when the disk is completely opaque to ionizing photons, $\eta = 0$,
which reduces the ionizing photon flux by less than a factor of two.

Estimates of $T_{\scriptscriptstyle h}$, $r_c$ and $n_c$ have come from 
three independent lines of inquiry. Absorption lines towards stars in the 
Magellanic Clouds have been used to infer coronal temperatures 
and gas densities; \eg Songaila (1981) finds 
$T_{\scriptscriptstyle h}\approx 0.05-0.3$
keV ($0.6-3.5 \times 10^6$ K) and $n_c\sim 3-7 \times 10^{-4}$
cm$^{-3}$. Alternatively, various groups have attempted to model the
Magellanic Stream as clumps moving through a hot, viscous fluid
(\qv Moore \& Davis 1994).  
An additional constraint is provided by pulsar
dispersion measures ($\Dm = 25-100$ cm$^{-3}$ pc) along sight lines to
the LMC/SMC system, and a halo globular cluster, NGC 5024 (Taylor,
Manchester \& Lyne 1993); these values imply electron densities of
order 10$^{-3}$ cm$^{-3}$. However, the \Dm\ column arises primarily
within the diffuse ionized gas layer in the Galactic disk (the
Reynolds layer), so that the true electron densities must be
considerably smaller.  Following Moore \& Davis (1994), we adopt $r_c
= 10$ kpc and n$_{-3} = 2$, which predicts a density of $6\times
10^{-5}$ cm$^{-3}$ at $R_o = 55$ kpc, and assume 
$T_{\scriptscriptstyle h}\approx 0.2\;{\rm
keV}\approx 2.3\times 10^6$ K, the virial temperature of the Galactic
halo (for an assumed circular velocity of 220 km s$^{-1}$). This
density and temperature distribution matches the emission measure and
X-ray luminosity of the diffuse component determined by Wang \& McCray
(1993). (See also Maloney \& Bland-Hawthorn 1998.) For this
temperature, $\xi_i\simeq 0.3-0.4\times10^{-15}$ cm$^{-3}$ s$^{-1}$
sr$^{-1}$, with only a weak dependence on metallicity. 

\psfig{file=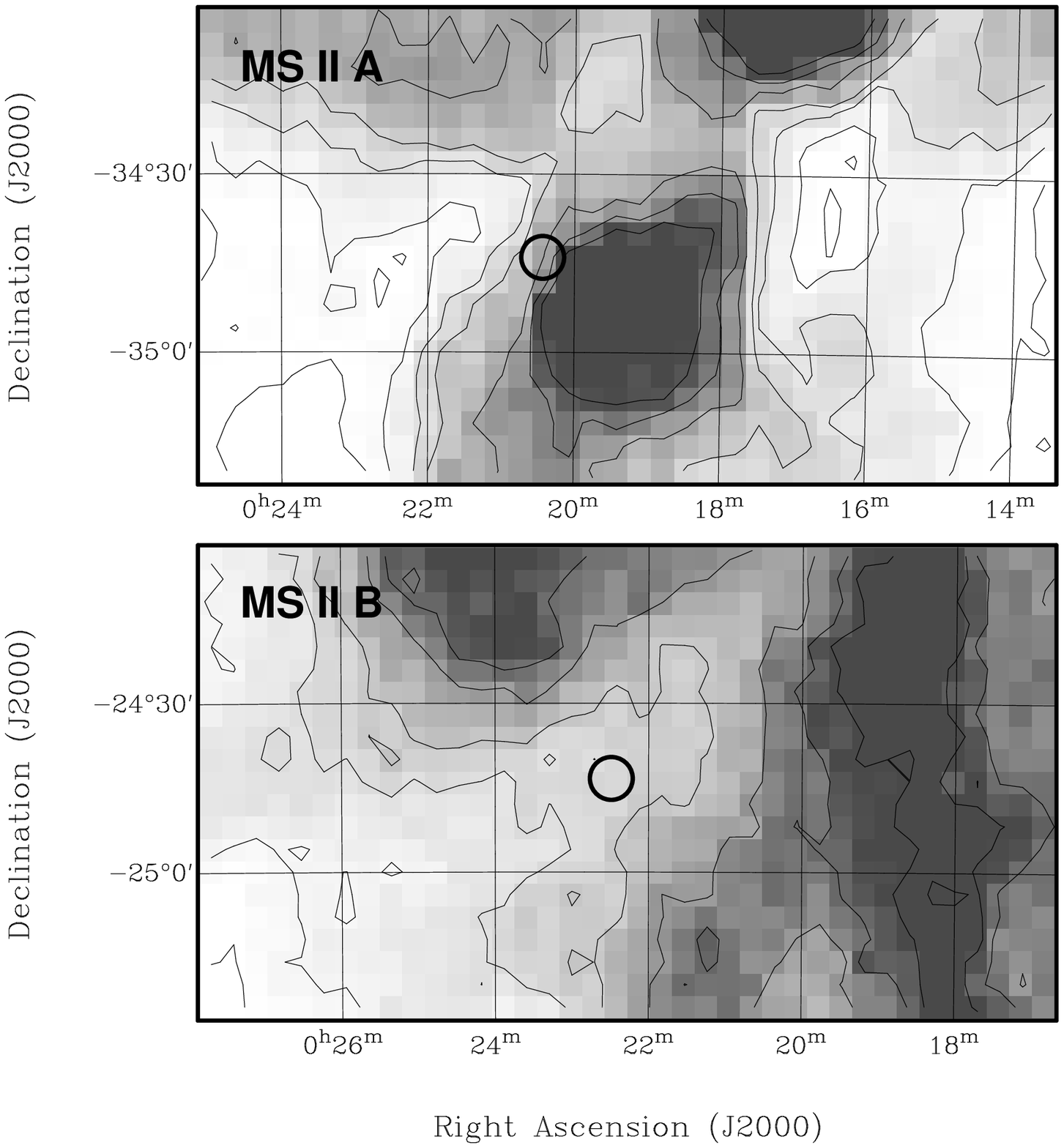,width=8cm}

\smallskip\noindent{\sl Fig.~3.}\ HI peak intensity map from the HIPASS 
survey of the Magellanic Stream (Putman\etal\ 1998) centered on two W$^2$ 
fields, where the W$^2$ beam size and pointings are shown by circles.
Strong \Ha\ detections fall in regions of high mean column density (\eg
MS~II~A); weak detections (\eg MS~II~B) and non-detections fall in regions 
of much lower column density. The contours are for brightness temperatures 
of 0.1, 0.2, 0.4, 0.6, 1, 1.4, 1.8 K.

\bigskip
These plasma constraints on the density and temperature imply that any 
contribution from a Galactic corona to the ionizing photon flux at the 
Stream is insignificant (eqn.~\ref{phoflux}).  For the known velocity 
of the Magellanic system ($\approx$215\kms; Lin, Jones \& Klemola 1995), 
and for volume-averaged, cloud densities of roughly 1 cm$^{-3}$, to produce 
substantial \Ha\ through shock ionization requires halo densities which 
are orders of magnitude higher than inferred from the pulsar dispersion 
measures.

\section{Photoionization of the Magellanic Stream}
\nobreak The Stream lies along a great arc which extends for more than
100$^{\circ}$ (\eg Mathewson, Cleary \& Murray 1974). Fig.~2
illustrates the relationship of the LMC to the Magellanic Stream above
the Galactic disk (Mathewson \& Ford 1984). The LMC lies close to the
$X$-$Z$ plane. We make the assumption that the Stream lies along
a circular orbit, close to the $X$-$Z$ plane, originating from the
Lagrangian point between the LMC and SMC, at a Galactocentric distance
of 55 kpc.\footnote{For the LMC, cepheids indicate $(m-M)_o = 
18.47\pm 0.15$, which implies 49.4$\pm$3.4 kpc (Feast \& Walker 1987); 
for the SMC, $(m-M)_o = 18.83\pm 0.15$, or 58.3$\pm$4.0 kpc (Feast
1988).} However, most computed orbits for the LMC-SMC system imply 
substantial ellipticity along the Stream (\eg Lin\etal\ 1995).

In order to derive \Ha\ emission measures, we assume an electron
temperature T$_e \approx 10^4$K, as expected for gas photoionized by
stellar sources; the Case B hydrogen recombination
coefficient is $\aB \approx 2.6 \times 10^{-13} (10^4/T_e)^{0.75}$
cm$^3$ s$^{-1}$. At these temperatures, collisional ionization
processes are negligible. In this case, for an optically thick slab,
the column recombination rate in equilibrium must equal the normally
incident ionizing photon flux, $\aB n_e N_i = \varphi_i$, where
\phiI\ is the rate at which Lyc photons arrive at the cloud surface
(\phoflux), $n_e$ is the electron density and $N_i$ is the column
density of ionized hydrogen and helium.

The emission measure is $\Em = \int n_e n_i\;dl =n_e
n_i L\ {\rm cm^{-6}\; pc}$ where $L$ is the thickness of the
ionized region.  The resulting emission measure for an ionizing flux
\phiI\ is then $\Em = 1.25\times 10^{-2} \varphi_4 \ {\rm
  cm^{-6}\; pc}$ ($= 4.5 \varphi_4 \ {\rm mR}$) where $\varphi_i =
10^4 \varphi_4$.  For an optically thin cloud in an isotropic
radiation field, the solid angle from which radiation is received is
$\Omega = 4\pi$, while for one-sided illumination, $\Omega=2\pi$.  For
our disk model, however, $J_\nu$ is anisotropic and $\Omega$ can be
considerably less than $2\pi$.

\psfig{file=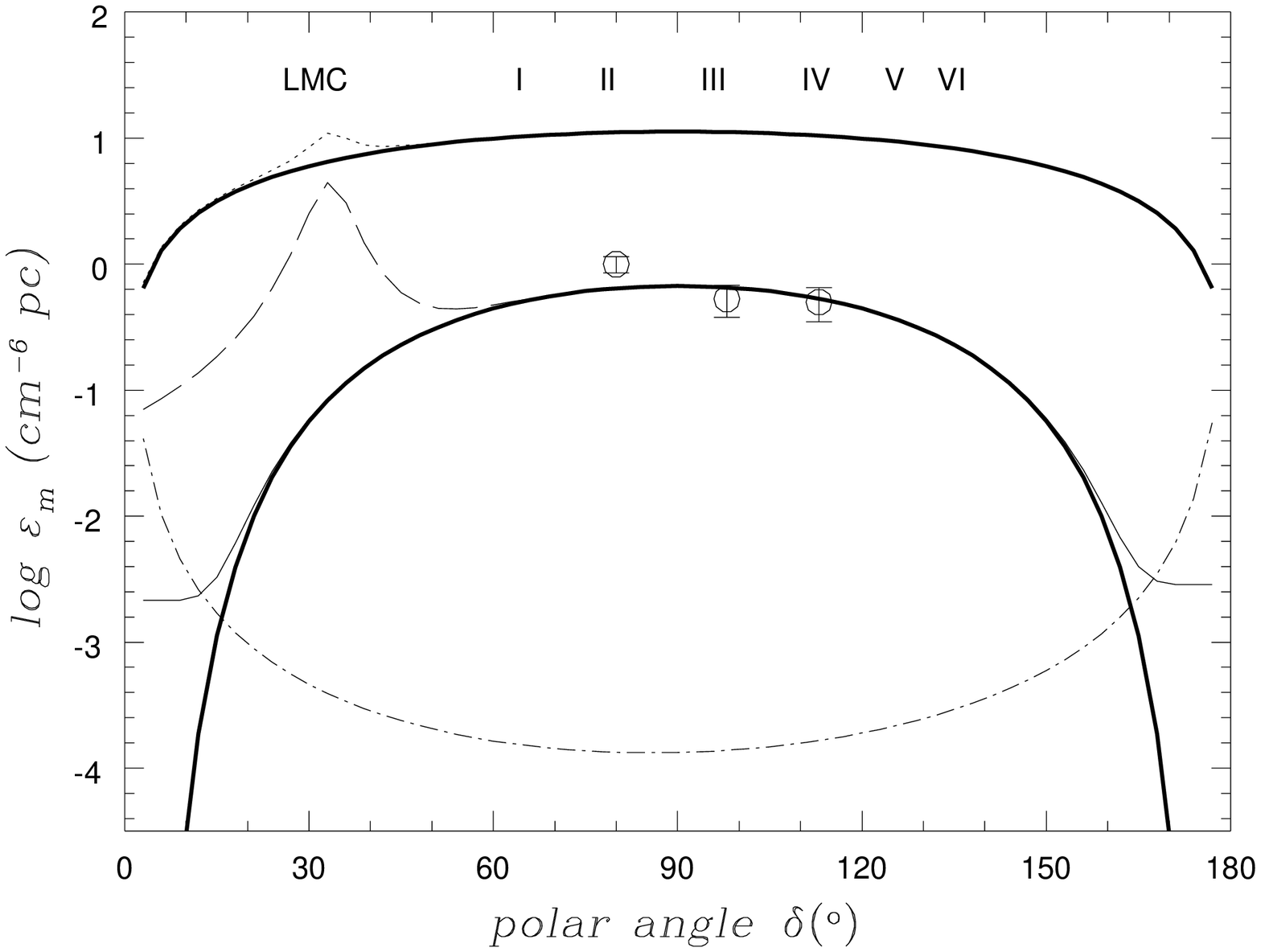,width=9cm}

\smallskip\noindent{\sl Fig.~4.}\ The predicted \Ha\ emission measure along the 
Stream as a function of $\delta$ (Fig.~2). The vertical axis has units of 
log(cm$^{-6}$ pc), equivalent to log(Rayleighs) after subtracting 
0.48.  The top curves assume an optically thin Galactic disk with (dashed 
line) and without (solid line) the LMC ionizing field. For the lower 
curves, the solid lines assume an opaque ($\tauLL = 2.8$) ionizing disk 
with (thin line) and without (thick line) a bremsstrahlung halo; the LMC 
contribution is shown by the long-dashed curve.
The dot-dash curve is $\Em$(\Ha) predicted for the upper side of the 
Stream due to the bremsstrahlung halo; the cosmic ionizing background is 
expected to dominate here. The open circles are the W$^2$ \Ha\ 
measurements.

\bigskip
The Stream \Ha\ detections are (1.1, 0.63, 0.60) \emunit\ or, equivalently, 
(370, 210, 200) mR\footnote{1 Rayleigh is 10$^6/4\pi$ phot cm$^{-2}$
s$^{-1}$ sr$^{-1}$ or 2.41$\times$10$^{-7}$ erg cm$^{-2}$ s$^{-1}$
sr$^{-1}$ at \Ha.}. Fig. 3 reveals that the Stream is highly clumped: 
the non-detections and weak detections fall in regions of low mean column
density ($\approx 1-3\times 10^{18}$ cm$^{-2}$), whereas the detections 
fall in regions of higher column density by at least an order of magnitude.
The near-constancy of the detected signal argues for a photoionization 
origin. We suspect that the projected gas is clumpy on all scales, and that
the detected regions are those in which the areal filling factor of 
neutral gas within the beam is essentially unity, so that all of the 
incident ionizing photons are absorbed by the gas. The close match between 
the \Ha\ and \HI\ kinematics also supports static photoionization. The
observed $\Em$ thus directly measures the ionizing photon flux 
incident on this portion of the Stream. We can therefore use the
\Ha\ detections to determine the escape fraction of 
ionizing photons from the Galaxy.

In Fig.~4, we show the predicted $\Em$ values along the Stream,
normalized to the \Ha\ detections. This requires a mean opacity
$\tauLL\approx 2.8\pm0.4$, corresponding to $\fesc \approx 6\%$. (Note
that for $\tauLL=0$, the Stream would be highly ionized.) The large
opacity of the disk elongates the surfaces of constant photon flux
density along the polar axis of the Galaxy (Bland-Hawthorn 1998). In 
this direction, the disk radiation field dominates over the cosmic
background (\Pcos\ $< 4\times 10^4\;\phoflux $: Vogel\etal\ 1995)
out to a distance of at least 200 kpc. We treat the LMC as a point 
source from which $5\times 10^{51}$ ionizing photons escape per
second (\eg Smith, Cornett \& Hill 1987). The Galactic radiation field 
dominates except within distances $\sim$10 kpc of the LMC.

\section{Discussion}
\nobreak
The conclusion that approximately 6\% of the ionizing photons
produced in the Galaxy escapes entirely from the disk and halo has
major implications for both the structure and energetics of the ISM
and the cosmic ionizing background. The uncertainty in this estimate 
is approximately 50\%, and is contributed equally by the
uncertainty in the ionizing flux incident on the Stream (\ie whether
the beam filling factor of high column density gas is actually unity
for any of the measurements), and by our knowledge of the total
ionizing photon luminosity of the Galaxy.

Reynolds (1990) has estimated that approximately 15\% of the ionizing
photons produced in the Galactic disk are needed to power the diffuse
ionized gas layer, \ie they must escape from the disk and reach heights 
of several hundred parsecs before absorption. Our inference that 
approximately 6\% of the ionizing photon luminosity escape the halo 
entirely lends substantial credence to the assertion of Miller \& Cox 
(1993) that the O star population of the Galaxy is responsible for the 
ionization and heating of the diffuse ionized gas layer.

A relatively intense, ionizing radiation field at large heights above the 
plane also has profound implications for the study of the high-velocity 
clouds (HVCs). Since their discovery more than 30 years ago, understanding 
of the nature and physical
characteristics of the HVCs has been severely hampered by the almost
complete absence of data on their distances (\qv Wakker \& van Woerden
1997). However, our determination of the ionizing radiation field in
the Galactic halo, normalized by the Stream detections, indicates that
it should be possible to derive \Ha\ distances to HVCs out to at least
$\sim$ 100 kpc (Bland-Hawthorn\etal\ 1998; \cf Tufte, Reynolds \& Haffner 
1998).

Finally, we note that if our value of \fesc\ for the Milky Way is 
typical, then galaxies dominate over AGN in producing the cosmic ionizing 
background at low redshift (Giallongo\etal\ 1997). 

\acknowledgments JBH thanks JILA and the University of Colorado for
their hospitality during the early stages of this work. The authors
gratefully acknowledge fruitful conversations with Nick Kaiser, Dennis 
Sciama (as always), Ken Freeman and Ron Reynolds. We were ably assisted 
by Gary da Costa, Peter Scheuer and Eric Smith. Ralph Sutherland assisted
generously with our emissivity calculations. We extend our warmest 
thanks Mary Putnam for providing Fig.~3 which was shown at a crucial moment 
during the Mt. Stromlo {\sl High Velocity Clouds} workshop earlier in the 
year.

\end{document}